\documentclass[prl,twocolumn,showpacs,preprintnumbers,superscriptaddress,amsmath,amssymb,floatfix]{revtex4-1}
\usepackage{graphicx}
\usepackage{amsmath}
\usepackage{epstopdf}
\usepackage{amsbsy}
\usepackage{xcolor}
\usepackage{dcolumn}
\usepackage{bm}

\newcommand{\bea}{\begin{eqnarray}}
\newcommand{\eea}{\end{eqnarray}}
\newcommand{\bt}{\textbf}

\newcommand{\phd}{\phantom{\dag}}
\newcommand{\ph}{\phantom{.}}

\newcommand{\noi}{\noindent}
\newcommand{\no}{\nonumber}

\begin{document}

\title{Disorder-Induced Electronic Nematicity}
\author{Daniel Steffensen}
\affiliation{Niels Bohr Institute, University of Copenhagen, Vibenshuset, Lyngbyvej 2, DK-2100 Copenhagen, Denmark}
\author{Panagiotis Kotetes}
\affiliation{Niels Bohr Institute, University of Copenhagen, Vibenshuset, Lyngbyvej 2, DK-2100 Copenhagen, Denmark}
\affiliation{CAS Key Laboratory of Theoretical Physics, Institute of Theoretical Physics, Chinese Academy of Sciences, Beijing 100190, China}
\author{Indranil Paul}
\affiliation{Laboratoire Mat\'{e}riaux et Ph\'{e}nom\`enes Quantiques, Universit\'{e} de Paris, CNRS,
F-75013, Paris, France}
\author{Brian M. Andersen}
\affiliation{Niels Bohr Institute, University of Copenhagen, Vibenshuset, Lyngbyvej 2, DK-2100 Copenhagen, Denmark}


\begin{abstract}
We expose the theoretical mechanisms underlying disorder-induced nematicity in systems exhi\-bi\-ting strong fluctuations or ordering in the nematic channel. Our analysis consists of a symmetry-based Ginzburg-Landau approach and associated microscopic calculations. We show that a single featureless point-like impurity induces nematicity locally, already above the critical nematic transition temperature. The persistence of fourfold rotational symmetry constrains the resulting disorder-induced nematicity to be inhomogeneous and spatially average to zero. Going beyond the single impurity case, we discuss the effects of finite disorder concentrations on the appearance of ne\-ma\-ti\-ci\-ty. We identify the conditions that allow disorder to enhance the nematic transition temperature, and we provide a concrete example. The presented theoretical results can explain a large series of recent expe\-ri\-men\-tal discoveries of disorder-induced nematic order in iron-based superconductors.
\end{abstract}


\maketitle

The study of electronic nematic quantum phases~\cite{NemReview} is becoming increasingly important in condensed matter systems due to a growing class of recently disco\-ve\-red ma\-te\-rials exhibiting nematic behavior, i.e. spontaneous ge\-ne\-ra\-tion of spatial anisotropy. Nematicity has been experimentally identified in a number of correlated quantum ma\-te\-rials~\cite{NemReview}, including quantum Hall states in higher Landau levels of 2D electron gases~\cite{QHEa,QHEb}, bilayer strontium ru\-the\-na\-tes~\cite{Ruthenates}, cuprate high-temperature superconductors (SCs)~\cite{Cuprates}, doped Bi$_2$Se$_3$ SCs~\cite{Bi2Se3a,Bi2Se3b,yonezawa}, Fe-based SCs (FeSCs)~\cite{tanatar,chu,ying,chu2,blomberg,ishida13,kuo14,yi11,kostin,zhao,nakajima,dusza,LGreene,kasahara} and, possibly, twisted bilayer graphene~\cite{Pashupathy18}. Thus, nematicity begins to establish as a universal electronic state of matter, which motivates further theoretical studies of its distinct properties.

Nematic phases are particularly prevalent in FeSCs, where experimental evidence for electronic nematicity comes from a wide
range of techni\-ques, including transport measurements~\cite{tanatar,chu,ying,chu2,blomberg,ishida13,kuo14},
angle-resolved photoemission spectroscopy~\cite{yi11}, scanning tunneling spectroscopy~\cite{kostin}, neutron scattering~\cite{zhao},
 light spectroscopy~\cite{nakajima,dusza,gallais2013},
Andreev-point-contact measurements~\cite{LGreene} and torque magnetometry~\cite{kasahara}.
In this case, the emergence of nematicity refers to the spontaneous brea\-king of fourfold (C$_4$) rotational symmetry.
Notably, the identification of the driving mechanism of nematicity in these systems is complicated, due to the coupling of spin,
orbital, and lattice degrees of freedom at tem\-pe\-ra\-tu\-res ($T$) below the tetragonal-to-orthorhombic structural
phase transition occurring at $T=T_s$~\cite{Fernandes14}. Particularly, the origin of nematicity in FeSe remains
controversial at present~\cite{AnnaAndreas}.

The growing ubiquity of nematic correlated electronic systems, that are scarcely free from impurities, calls for resolving the influence of disorder on the emergence of nematicity. In fact, the strong relevance of di\-sor\-der to the nematic ordering is also supported from a notable number of experiments detecting local C$_4$-symmetry breaking around impurities~\cite{chuang10,song11,zhou11,grothe12,allan13,Rosenthal14,kostin}. While some of these results may be attributable to, for instance, residual sample strain which explicitly breaks the C$_4$ symmetry locally~\cite{Ren15,Rosenthal14,Baek16}, or the presence of stripe-ordered antiferromagnetism~\cite{chuang10}, the possible pinning of nematic fluctuations due to the pre\-sen\-ce of disorder appears as a pro\-mi\-sing and, at the moment, poorly-explored alternative~\cite{chen09,Navarro_cigar,kontani12,Gastiasoro14,YWang15}. Even more, there are strong indications for disorder-pinned static local nematicity in the bulk te\-tra\-go\-nal phase, i.e. {\it above} $T_s$~\cite{Kasahara12,Iye15,Zhou16,Wiecki17,Wang17,Toyoda18}. For exam\-ple, two recent NMR experiments on FeSe~\cite{Wiecki17,Wang17}, found a clear splitting and broa\-de\-ning of the NMR lineshape above $T_s$. The pre\-sen\-ce of short-range nematic order above the bulk $T_s$ in FeSe has also been inferred from ARPES and optical-pump conductivity measurements~\cite{Zhang15,Luo17}. Finally, two very recent pair distribution function (PDF) measurements of FeSe found clear evidence of pronounced local orthorhombicity at the length scale of a few nm well above $T_s$~\cite{koch19,frandsen19}, thus providing additional evidence for disorder-induced local nematicity in these systems.

In this Letter, we perform a detailed theo\-re\-ti\-cal study of the role of disorder in systems with D$_{\rm 4h}$ point-group symmetry, which additionally feature strong fluctuations or ordering in the nematic channel. The emergence of nematicity is reflected in a non-zero field $N$, which transforms according to the B$_{\rm 1g}$ irreducible representation (IR) of D$_{\rm 4h}$. We mainly focus on $T$ above the respective $T_{\rm nem}$ (same as $T_s$), at which, the spontaneous thermodynamic C$_4$-symmetry breaking takes place. By employing both phenomenological and microscopic approaches, we address the following three questions: 1) Under what circumstances can disorder ge\-ne\-ra\-te nematicity locally? 2) What is the spatial profile of the resulting nematic-defect structure? 3) How do finite disorder concentrations influence the nematic transition?

Our main results can be summarized as follows. For $T>T_{\rm nem}$, \bt{(i)} an impurity potential of arbitrary strength
with a spatial profile which respects the C$_4$ symmetry, generates a local nematic field $N(\bm{r})$ with a
spatial profile belon\-ging to the B$_{\rm 1g}$ IR. By transferring to a polar coordinate system $(x,y)\mapsto(r,\phi)$,
this yields the spatial profile $N(r,\phi)\propto \cos(2\phi)$. \bt{(ii)} This further implies, that, a potential
with a C$_4$-symmetric profile does not induce net nematicity, i.e., $\int d\bm{r} N(\bm{r})=0$, but local probes may still detect evidence of C$_4$ symmetry breaking. \bt{(iii)} However,
we show that such a potential can still drive a nematic transition already at $T>T_{\rm nem}$, since it mo\-di\-fies
the Stoner criterion for the nematic instability. \bt{(iv)} A C$_4$-symmetry-breaking impurity potential can induce
nonzero net nematicity and, thus, stabilize long-range nematic order. For $T<T_{\rm nem}$, an impurity potential with a
spatial profile which respects C$_4$ symmetry modifies the bulk nematicity ($N_{\rm B}$) locally, and results in an
inhomogeneous nematic field $N(\bm{r})=N_{\rm B}+\delta N(\bm{r})$, with a $\delta N(r,\phi)$ which is generally not
proportional to $\cos(2\phi)$.

We first examine the implications of disorder within a continuum Ginzburg-Landau (GL) approach, that allows exposing generic features of the induced nematic field, i.e. independent of the origin of the electronic nematicity. In fact, our GL results also apply to situations where the nematic field originates from the spontaneous mixing of superconducting order parameters belon\-ging to the A$_{\rm 1g}$ and B$_{\rm 1g}$ IRs~\cite{Livanas,Fernandes-Millis}. However, there, one has to further include the possible influence of disorder on the pairing.

The free energy density ${\cal F}(\bm{r})$ is a functional of $N(\bm{r})$ and the disorder potential $V(\bm{r})$. Its invariance under D$_{\rm 4h}$ point group operations and time reversal, leads to:
\bea
{\cal F}(\bm{r})&=&\alpha(T-T_{\rm nem})[N(\bm{r})]^2/2+\beta[N(\bm{r})]^4/4\no\\
&+&c[\bm{\nabla}N(\bm{r})]^2/2+gN(\bm{r})\left(\partial_x^2-\partial_y^2\right)V(\bm{r}),\qquad\label{eq:Free}
\eea

\noi with $\alpha,\beta,c>0$. Here, we restricted to the \textit{lowest-order} possible coupling between $V(\bm{r})$ and $N(\bm{r})$. Later on, we consider effects of higher-order terms. For further details on the GL approach, we refer to the Supplementary Material (SM)~\cite{Suppl}. From Eq.~\eqref{eq:Free}, one observes that the nematic field couples to the second derivatives of the disorder potential and, thus, to a particular li\-near combination of the electric field gradients (EFGs). The nematic field is proportional to the quadrupolar electronic charge density defined as $Q_{x^2-y^2}(\bm{r})=\big(x^2-y^2\big)\rho(\bm{r})$, which transforms accor\-ding to the B$_{\rm 1g}$ IR of D$_{\rm 4h}$, i.e. similar to $N(\bm{r})$. In the above, $\rho(\bm{r})$ denotes the electric charge density, which belongs to the trivial (A$_{\rm 1g}$) IR of D$_{\rm 4h}$. The appearance of a nonzero $N(\bm{r})$, solely due to the presence of disorder, is a consequence of the broken translational invariance, and can be viewed as a result of linear response, since the EFG $\left(\partial_x^2-\partial_y^2\right)V(\bm{r})$ acts as a quadrupolar source field, which leads to a nonzero and \textit{necessarily} inhomogeneous $Q_{x^2-y^2}(\bm{r})$ and thus $N(\bm{r})$.

For the remainder, we consider $T>T_{\rm nem}$ (unless explicitly stated), which implies that the system resides in the C$_4$-symmetric phase in the absence of disorder. In this case, we can drop the quartic nematic term, since $N(\bm{r})$ is generally small. Thus, for $T>T_{\rm nem}$, the Euler-Lagrange equation of motion (EOM) for Eq.~\eqref{eq:Free} reads:
\bea
\big[\alpha(T-T_{\rm nem})-c\bm{\nabla}^2\big]N(\bm{r})=-g\left(\partial_x^2-\partial_y^2\right)V(\bm{r})\,.\quad\label{eq:NematicProfileReal}
\eea

\begin{figure}[t!]
\centering
\includegraphics[width=1\columnwidth]{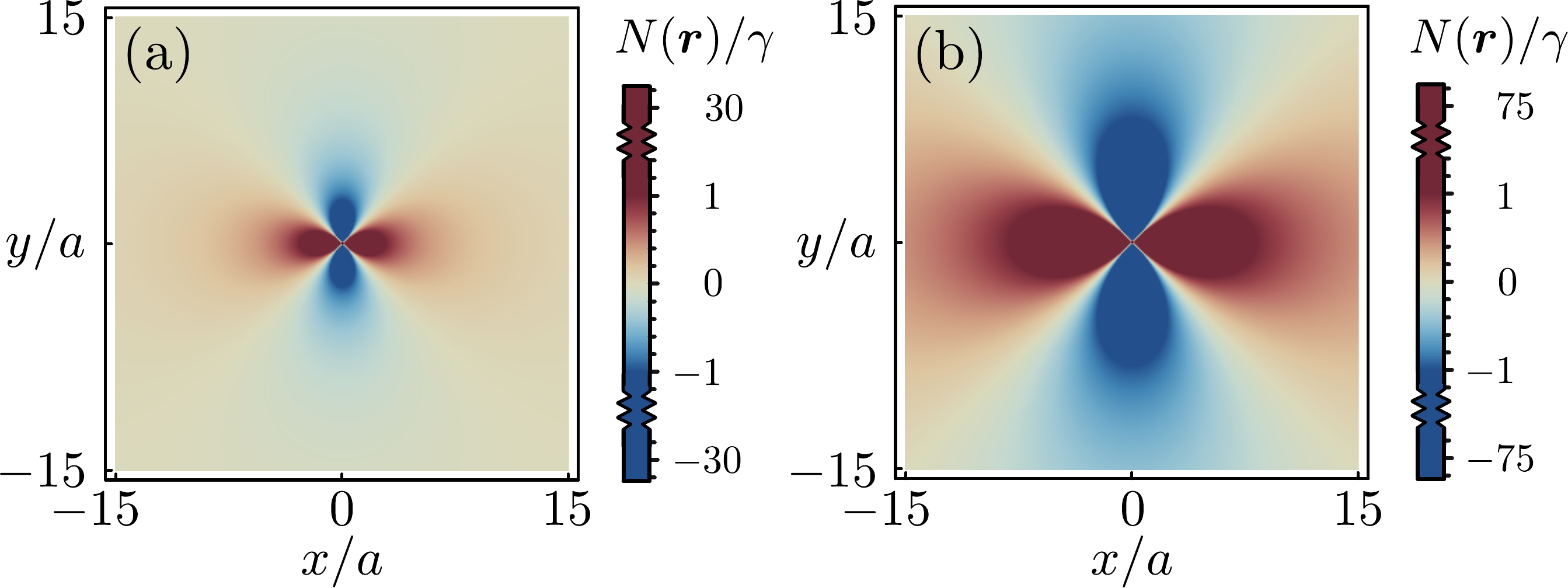}
\protect\caption{(a) Nematic order parameter $N(\bm{r})$ at $T \gg T_{\rm nem}$, where $\xi_{\rm nem} \sim 5\, a$. (b) Same as in (a), but with $T \gtrsim T_{\rm nem}$ resulting in a larger nematic coherence length $\xi_{\rm nem}\sim15\,a$. The figures were obtained using Eq.~\eqref{eq:NematicProfile} with a convenient impurity profile of the form $V(\bm{r})=V/|\bm{r}|$, without loss of generality. We introduced $\gamma=-\pi gV/(2\xi_{\rm nem})$ and used $c=1$.}
\label{fig:Figure1}
\end{figure}

\noi The above EOM provides the proportionality relation between the EFG and the resulting nematic field, i.e.:
\bea
N(\bm{r})=\int \frac{d\bm{q}}{(2\pi)^2}e^{i\bm{q}\cdot\bm{r}}\ph\frac{g}{c}\frac{\big(q_x^2-q_y^2\big)V(\bm{q})}{\bm{q}^2+\xi_{\rm nem}^{-2}}\,,\label{eq:NematicProfile}
\eea

\noi where we introduced the nematic coherence length in the tetragonal phase $\xi_{\rm nem}^{-1}=\sqrt{\alpha(T-T_{\rm nem})/c}$.

For a C$_4$-symmetric impurity potential we integrate the angular part of the rhs in Eq.~\eqref{eq:NematicProfile}, and
find the earlier-announced spatial profile $N(r,\phi)\propto \cos(2\phi)$. This profile decays away from the impurity
within a range given by  $\xi_{\rm nem}$, and  this length scale
diverges as $T\rightarrow T_{\rm nem}$. Both results are depicted in Fig.~\ref{fig:Figure1}. The angular dependence transforms exac\-tly according to the B$_{\rm 1g}$ IR of D$_{\rm 4h}$. This constraint on the spatial profile of $N(\bm{r})$ is a directly consequence of the featureless (A$_{\rm 1g}$) nature of the disor\-der potential $V$ itself. Thus, the net electronic nemati\-ci\-ty and quadrupolar charge are zero, since:
\bea
\int d\bm{r}\ph N(\bm{r})=N(\bm{q}=\bm{0})\propto\int_0^{2\pi}d\phi\ph N(r,\phi)=0\,.\label{eq:NematicZero}
\eea

\noi Nonetheless, probes like NMR and PDF pick up a signal from atoms in the lobes of the induced $N(\bm{r})$, and do therefore detect clear evidence for local nematicity and orthorhombicity even though global effects are absent.

Equation~\eqref{eq:NematicZero} also reveals that the linear coupling of the nematic field to the EFG cannot stabilize a net thermodynamic nematicity which emerges when $N(\bm{q}=\bm{0})\neq0$. Therefore, the quadrupolar coupling can neither preempt nor smear out the bulk nematic phase transition. A nonzero $N(\bm{q}=\bm{0})$ can, however, be induced when the spatial profile of the disorder potential expli\-ci\-tly breaks C$_4$ symmetry. This can be seen by inclu\-ding higher-order couplings in the GL free energy (see also SM~\cite{Suppl}):
\bea
\delta{\cal F}(\bm{r})=-\bigg\{g'V(\bm{r})+g''\big[V(\bm{r})\big]^2\bigg\}\big[N(\bm{r})\big]^2/2\,.\label{eq:HOT}
\eea

\noi The above terms provide couplings between $V(\bm{q}\neq\bm{0})$ and $N(\bm{q}=\bm{0})$, as well as the $N(\bm{q}\neq\bm{0})$ nematic-field components. These couplings are essential to describe a disorder-driven preemptive nematic transition above $T_{\rm nem}$, as well as the emergence of net nematicity when the potential breaks C$_4$ symmetry. To demonstrate both aspects, we derive the modified EOM for the $N(\bm{q}=\bm{0})$ component after adding the contribution of Eq.~\eqref{eq:HOT} to the free energy of Eq.~\eqref{eq:Free}. We find the following EOM:
\bea
&&\alpha(T-T_{\rm nem})N(\bm{q}=\bm{0})=g'\int d\bm{p}\ph V(\bm{p})N(-\bm{p})\no\\
&&\quad+g''\iint d\bm{p}d\bm{p}'\ph V(\bm{p}')V(\bm{p}-\bm{p}')N(-\bm{p})\,.
\eea

\noi Thus, a nonzero $N(\bm{q}=\bm{0})$ can only emerge when components with $\bm{q}\neq\bm{0}$ are already nonzero. By assuming that the potential is weak, the $N(\bm{q}\neq\bm{0})$ components remain given by the Fourier transform of Eq.~\eqref{eq:NematicProfileReal}. Therefore, we obtain the following equation up to se\-cond order in $V(\bm{r})$:
\bea
\left[\alpha(T-T_{\rm nem})-g''\int d\bm{p}\ph|V(\bm{p})|^2\right]N(\bm{q}=\bm{0})\no\\
=\frac{gg'}{c}\int d\bm{p}\ph\frac{\big(p_x^2-p_y^2\big)|V(\bm{p})|^2}{\bm{p}^2+\xi_{\rm nem}^{-2}}\,.\quad\label{eq:NetNematicityEOM}
\eea

Eq.~\eqref{eq:NetNematicityEOM} implies that a C$_4$-symmetric configuration of impurities cannot source a homogeneous component for the nematic field, since the rhs is zero. As we prove in SM~\cite{Suppl}, this holds even after including all the symmetry-allowed higher-order GL terms. In fact, this result is also recovered in the case of a large number of randomly-distributed and uncorrelated impurities, in which si\-tua\-tion, translational and rotational symmetries are preserved on average. Thus, a C$_4$-symmetric disorder potential solely modifies the nematic Stoner criterion, i.e.:
\bea
T_{\rm nem}^{\rm imp}=T_{\rm nem}+\frac{g''}{\alpha}\int d\bm{p}\ph|V(\bm{p})|^2\,.\label{eq:ModTemp}
\eea

\noi Depending on the microscopic details which control the sign of the coupling constant $g''$, the nematic transition temperature can be enhanced. Note, however, that such an enhancement tends to zero in the thermodynamic limit, unless a critical density of impurities $n_{\rm imp}$ is present. This is because the $g''$ coefficient is inversely proportional to the system size. Interestingly, a detailed transport study with controlled disorder by electron irradiation found cases where the critical nematic transition temperature increased slightly with disorder~\cite{Timmons2019}.

Before proceeding, we point out that the first term of Eq.~\eqref{eq:HOT} also allows to describe the induced net ne\-ma\-ti\-ci\-ty when the disorder potential breaks C$_4$ symmetry. To exem\-pli\-fy this, we consider a dimer impurity potential $V(\bm{r})=V\left[\delta(\bm{r}-\hat{\bm{x}})+\delta(\bm{r}+\hat{\bm{x}})\right]$, which yields $V(\bm{p})=V(\cos p_x+\cos p_y)+V(\cos p_x-\cos p_y)$, for a lattice constant $a = 1$. The brea\-king of C$_4$ symmetry is ensured by the combined pre\-sen\-ce of the A$_{\rm 1g}$ and B$_{\rm 1g}$ IRs. In general, a nonzero $N(\bm{q}=\bm{0})$ arises whenever $|V(\bm{q})|^2$ contains at least one B$_{\rm 1g}$ term.

To support the above GL findings, we employ a microscopic tight-binding model of electrons coupled to disorder. This analysis not only verifies the above GL results, but more importantly, allows to uncover further microscopic details which control the emergence of nematicity. In the absence of disor\-der, the electrons are described by the dispersion $\varepsilon_{\bm{k}}=-2t(\cos k_x+\cos k_y)-\mu$. The spin degree of freedom is neglected throughout this work, since it merely introduces a factor of 2 in all thermodynamic ave\-ra\-ges. We assume that the electrons feel an attractive effective interaction in the Pomeranchuk nematic channel as in Ref.~\onlinecite{Gallais}, which, after mean-field decoupling, yields the nematic order parameter (for details see SM~\cite{Suppl}):
\bea
N_{\bm{R}}=-V_{\rm nem}\sum_{\bm{\delta}}f_{\bm{\delta}}\big<c^{\dag}_{\bm{R}+\bm{\delta}}c_{\bm{R}}+c^{\dag}_{\bm{R}}c_{\bm{R}+\bm{\delta}}\big>\,,
\label{eq:SelfConsistency}
\eea

\noi i.e. the lattice analog of $N(\bm{r})$. This introduces a local or global C$_4$-breaking to the electron-hopping matrix ele\-ments. In the above, $c_{\bm{R}}$ denotes the annihilation ope\-ra\-tor of an electron at position $\bm{R}=(n,m)$ of the lattice, with $n,m\in\mathbb{Z}$. In addition, $\hat{\bm{x}}$ ($\hat{\bm{y}}$) corresponds to the unit vector in the $x$ ($y$) direction. The nematic form factor is nonzero for $\bm{\delta}=\pm\hat{\bm{x}},\pm\hat{\bm{y}}$, and reads $f_{\pm\hat{\bm{x}}}=-f_{\pm\hat{\bm{y}}}=1/4$. Disorder is considered in the form of point-like identical impurities. The total mean-field Hamiltonian becomes:
\bea
\widehat{{\cal H}}&=&\sum_{\bm{R},\,\bm{\delta}}\Big(N_{\bm{R}}f_{\bm{\delta}}-t/2\Big)\Big(c^{\dagger}_{\bm{R}+\bm{\delta}}c_{\bm{R}}+{\rm h.c.}\Big)\no\\
&+&\sum_{\bm{R}}\left( V_{\bm{R}}-\mu \right)c^{\dag}_{\bm{R}}c_{\bm{R}}\,.
\label{eq:Hfull}
\eea

\begin{figure}[t!]
\centering
\includegraphics[width=1\columnwidth]{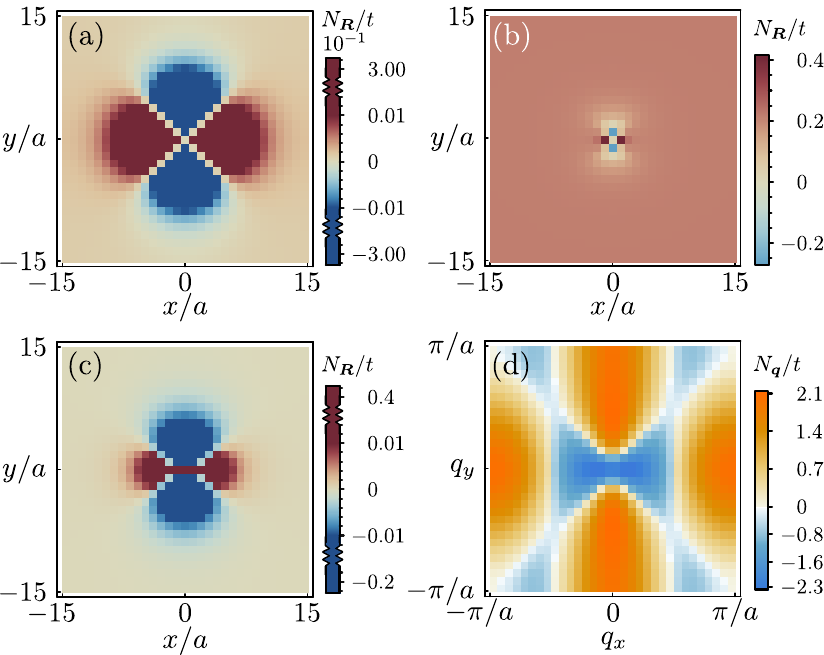}
\protect\caption{Numerically-obtained nematic order parameter using the microscopic model of Eq.~\eqref{eq:Hfull}: (a) displays the local nematic order pinned by a single impurity at $\bm{R}=\bm{0}$ for $T=0.8\,t$. For the given set of parameters, the Stoner criterion is fulfilled for $T\sim 0.78\,t$. (b) Same as in (a), but in the bulk nematic phase ($T=0.76\,t$). (c) Induced nematic order in the presence of a dimer impurity potential, $V_{\bm{R}}=V(\delta_{\bm{R},\hat{\bm{x}}}+\delta_{\bm{R},-\hat{\bm{x}}})$, and (d) its discrete Fourier transform ($T=0.8\,t$). From panel (d), one clearly sees that the breaking of C$_4$ symmetry indeed induces $N_{\bm{q}=\bm{0}}\neq0$. All the figures were obtained using: $V=5\,t$, $\mathcal{N}_x=\mathcal{N}_y=31$, $V_{\rm nem}=4\,t$, $k_{\rm B}=1$ and $\langle n\rangle=0.53$.}
\label{fig:Figure2}
\end{figure}

For a single delta-function impurity potential $V_{\bm{R}}=V\delta_{\bm{R},\bm{0}}$, we evaluate the nematic order parameter in Eq.~\eqref{eq:SelfConsistency} self-consistently for a fixed electron density $\langle n \rangle$ (see SM~\cite{Suppl}). The resulting nematic order is displayed in Fig.~\ref{fig:Figure2}(a), and possesses the same spatial profile as those shown in Fig.~\ref{fig:Figure1}. In the case of a dimer impurity potential $V_{\bm{R}}=V(\delta_{\bm{R},\hat{\bm{x}}}+\delta_{\bm{R},-\hat{\bm{x}}})$, which explicitly breaks C$_4$ symmetry, we obtain the profile shown in Fig.~\ref{fig:Figure2}(c). Its Fourier transform, see Fig.~\ref{fig:Figure2}(d), exhibits $N_{\bm{q}=\bm{0}}\neq0$, which ori\-gi\-na\-tes from the rhs of Eq.~\eqref{eq:NetNematicityEOM}. We stress that, the fact that the induced clover pattern in Fig.~\ref{fig:Figure2}(a) is directly sourced by the EFG, makes it distinct from other microscopic studies of impurity-induced local order~\cite{andersen07,romer2012,Navarro_LiFeAs}. There, the impurity potential induces a spontaneous symmetry breaking locally, by means of a local fulfillment of the Stoner criterion, i.e. analogously to Eq.~\eqref{eq:ModTemp}.

We proceed by studying the effects of a single impurity for $T<T_{\rm nem}$, where the system resides in the bulk phase with a value $N_{\rm B}$ for the nematic order parameter. In this case, the order parameter assumes the form $N(\bm{r})=N_{\rm B}+\delta N(\bm{r})$, where $\delta N(\bm{r})$ incorporates the spatial variation of the nematic order pa\-ra\-me\-ter near the impurity. For a weak impurity potential, we expand the EOM stemming from Eq.~\eqref{eq:Free} up to linear order in $\delta N(\bm{r})$ (see SM~\cite{Suppl}). We find that $\delta N(\bm{r})$ possesses the spatial profile of Eq.~\eqref{eq:NematicProfile}, with the difference that the cohe\-ren\-ce length is now given by $\xi_{\rm nem}^{-1}=\sqrt{2 \alpha(T_{\rm nem}-T)/c}$ due to an additional contribution of the quartic term which has to be taken into account for $T<T_{\rm nem}$. From a microscopic calculation, we obtain the spatial profile for the nematic order pa\-ra\-me\-ter which is shown in Fig.~\ref{fig:Figure2}(b), exhi\-bi\-ting an anisotropic local structure which is slightly elongated along the $y$ direction. To lowest order in $V(\bm{r})$, this asymmetry found via the microscopic model can be reproduced in the GL theo\-ry by including the first term of Eq.~\eqref{eq:HOT}. The presence of this term yields $\delta N(\bm{r})\propto f(r)\cos(2\phi)+h(r)N_{\rm B}$, where $f(r)$ and $h(r)$ are decaying functions of the radial coordinate, transfor\-ming according to the A$_{\rm 1g}$ IR. Note that additional higher order terms, e.g. $\propto V(\bm{r})(\partial_x^2+\partial_y ^2)[N(\bm{r})]^2$, can further contribute to this anisotropy by modifying $h(r)$. In ge\-ne\-ral, we find that depending on the sign of the impurity potential, point-like disorder at $T<T_{\rm nem}$ may either locally enhance or decrease the nematic order.

Finally, we verify the possibility of disorder-enhanced $T_{\rm nem}$ within the microscopic model. We assume random and dilute disorder of density $n_{\rm imp}$, that on average preserves the C$_4$ symmetry. Within the 1$^{\rm st}$ order Born approximation~\cite{Bruus-Flensberg}, the quasiparticle lifetime is:
\bea
\frac{1}{\tau_{\bm{k}}}=2\pi n_{\rm imp}V^2\frac{1}{\cal N}\sum_{\bm{p}}\delta_{\varepsilon_{\bm{p}},\varepsilon_{\bm{k}}}\,.\label{eq:Lifetime}
\eea

\noi By use of Eq.~\eqref{eq:Lifetime}, we can eva\-lua\-te the microscopic coefficients which enter the mo\-di\-fied Stoner criterion of Eq.~\eqref{eq:ModTemp}, brought about by the impurities. Starting from Eq.~\eqref{eq:SelfConsistency}, we find that the self-consistency equation for the $\bm{q}=\bm{0}$ component of the nematic mean-field order parameter, corresponding to net nematicity $N\equiv\sum_{\bm{R}}N_{\bm{R}}/{\cal N}=N_{\bm{q}=\bm{0}}/{\cal N}$, reads
\bea
N=-V_{\rm nem}\frac{1}{\cal N}\sum_{\bm{k}}f_{\bm{k}}\int_{-\infty}^{+\infty}\frac{d\varepsilon}{2\pi}
\frac{n_F(\varepsilon+\varepsilon_{\bm{k}}+Nf_{\bm{k}})}{\varepsilon^2+1/(2\tau_{\bm{k}})^2}\frac{1}{\tau_{\bm{k}}}\,,\no
\eea

\noi with $f_{\bm{k}}=\cos k_x-\cos k_y$ and $n_F(\varepsilon)$ the Fermi-Dirac distribution function. Linearizing the rhs with respect to $N$, yields the mo\-di\-fied Stoner criterion
\bea
\frac{1}{V_{\rm nem}}=-\frac{1}{\cal N}\sum_{\bm{k}}f_{\bm{k}}^2\int_{-\infty}^{+\infty}\frac{d\varepsilon}{2\pi}\frac{n_F'(\varepsilon+\varepsilon_{\bm{k}})}{\varepsilon^2+1/(2\tau_{\bm{k}})^2}\frac{1}{\tau_{\bm{k}}}\,.
\label{eq:mod_stoner}
\eea

\begin{figure}[t!]
\centering
\includegraphics[width=1\columnwidth]{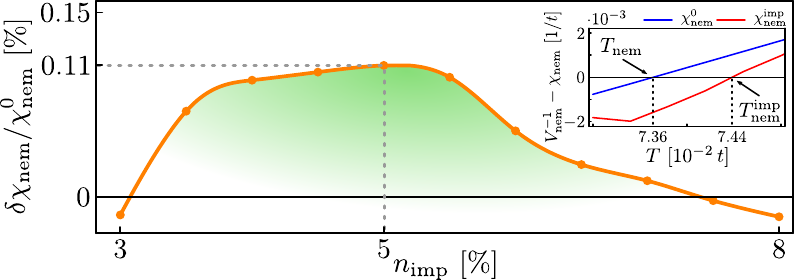}
\protect\caption{Relative disorder-induced modification of the nematic susceptibility $\delta\chi_{\rm nem}/\chi_{\rm nem}^0=(\chi_{\rm nem}^{\rm imp}-\chi_{\rm nem}^{0})/\chi_{\rm nem}^{0}$ versus the di\-sor\-der concentration $n_{\rm imp}$. Here, $\chi_{\rm nem}^0$ is the rhs of Eq.~\eqref{eq:mod_stoner} in the absence of disorder $\tau_{\bm{k}}\rightarrow\infty$. The inset shows that the $T_{\rm nem}$ increases by approximately $1\%$ for $n_{\rm imp}\approx5\%$, due to the disorder-modified Stoner criterion. Parameters: $V_{\rm nem} = 1.584\,t$, $\langle n\rangle=0.53$, $\mathcal{N}_{x}=\mathcal{N}_y=201$, $T=0.075\,t$ and $V=5\,t$.}
\label{fig:Figure3}
\end{figure}

In the absence of disorder, i.e. $\tau_{\bm{k}}\rightarrow\infty$, the integration yields the derivative of the Fermi-Dirac distribution $n_F'(\varepsilon_{\bm{k}})$. However, for finite $\tau_{\bm{k}}$, each $\bm{k}$ state is broa\-de\-ned, and the density of states (DOS) for the $\bm{k}$ points mainly contributing to the nematic instability may be enhanced. To explore this effect, we numerically calculate the nematic susceptibility $\chi_{\rm nem}^{\rm imp}$ in the presence of disorder, which is identified with the rhs of Eq.~\eqref{eq:mod_stoner}. Fig.~\ref{fig:Figure3} shows how this quantity changes versus the disorder concentration $n_{\rm imp}$, relative to the disorder-free case. For an impurity density of $n_{\rm imp}\approx 5\%$, we obtain the maximal relative enhancement of $\chi_{\rm nem}^{\rm imp}$ leading to a correspon\-ding small enhancement of $T_{\rm nem}$. It is tempting to assign the similar small enhancement of the nematic transition temperature measured experimentally in Ref.~\onlinecite{Timmons2019} to the effect demonstrated in Fig.~\ref{fig:Figure3}.

The origin of the enhancement effect shown in Fig.~\ref{fig:Figure3} is the presence of a nearby van Hove singularity whose spectral weight can be utilized to boost $\chi_{\rm nem}^{\rm imp}$ in the presence of finite $\tau_{\bm{k}}$. Without favorable DOS conditions, di\-sor\-der generally suppresses the nematic susceptibility and hence also $T_{\rm nem}$. Such a suppression tendency has been pre\-vious\-ly found in Ref.~\onlinecite{Schofield} and is also demonstrated in our SM~\cite{Suppl}. Further, we remark that even in the disorder-free case, the presence of a van Hove singularity is pivotal for the stabilization of an electron nematic phase of the Pomeranchuk type. For more details see Refs.~\cite{Kee,Yamase}.

In summary, we have elucidated the coupling of nematicity to disorder from both a phenomenological GL approach and microscopic calculations. Importantly, disorder is always locally relevant for inducing nemati\-ci\-ty since the EFG $\left(\partial_x^2-\partial_y^2\right)V(\bm{r})$  directly acts as a quadrupolar source field for nematicity. This explains the detection of local nematicity/orthorhombicity in experimental probes sensitive to different atomic environments within materials. At the global scale, however, disorder does not generally generate long-range nematicity at $T>T_{\rm nem}$ where the system remains tetragonal. Finite disorder concentrations may, however, under favorable circumstances enhance nematic order.

We thank Andreas Kreisel and Karsten Flensberg for useful discussions. D.~S. and B.~M.~A. acknowledge financial support from
the Carlsberg Foundation. P.~K. and B.~M.~A. acknow\-ledge support from the Independent Research Fund Denmark grant
number DFF-6108-00096. I.~P. is supported by ANR grant ``IRONIC'' (ANR-15-CE30-0025).


\pagebreak
\def\v#1{{\bf #1}}
\begin{widetext}
\begin{center}
\textbf{\large Supplementary Material:\\ "Disorder-Induced Electronic Nematicity"}
\end{center}
\end{widetext}

\section{Ginzburg-Landau Theory: Phenomenological Analysis}

Eq.~(1) of the main text is obtained by demanding that the free energy density is a real functional transforming according to the trivial irreducible representation (IR) of the ensuing point group. Here, we assume a system with tetragonal and inversion symmetries present, which is described by the D$_{\rm 4h}$ point group symmetry. The free energy density transforms according to the A$_{\rm 1g}$ IR of D$_{\rm 4h}$ and is here also assumed invariant under time reversal. 

\subsection{Equation of Motion}

The equation describing the nematic field is found via the Euler-Lagrange equation of motion (EOM):
\bea
\frac{\partial{\cal F}}{\partial N}=\partial_x\frac{\partial{\cal F}}{\partial(\partial_xN)}+\partial_y\frac{\partial{\cal F}}{\partial(\partial_yN)}
\eea

\noi and reads:
\bea
&&\big[\alpha(T-T_{\rm nem})-c\bm{\nabla}^2\big]N(\bm{r})+\beta[N(\bm{r})]^3\quad\no\\
&&\qquad\qquad=-g\left(\partial_x^2-\partial_y^2\right)V(\bm{r})\,.\phd\quad\qquad\label{eq:EOMfull}
\eea

\noi From the above, one notes that if the potential $V(\bm{r})$ is homogeneous, i.e. $V(\bm{r})=V$, the EOM includes only derivatives of $N$ and no other spatially-dependent functions or source terms. Thus, for an infinite (bulk) system $N(\bm{r})=N$. When $T>T_{\rm nem}$, the appearance of ne\-ma\-ti\-ci\-ty is disfavored and, thus, $N=0$ in the bulk. In contrast, the presence of an inhomogeneous potential functions as a source of nematicity and allows for non-zero inhomogeneous solutions of $N(\bm{r})$. 

\subsection{Case Study: Single Impurity for $\bm{T>T_{\rm nem}}$} 

Above $T_{\rm nem}$, we drop the cubic term in the EOM in Eq.~\eqref{eq:EOMfull} above, and obtain Eq.~(2) of the manuscript. For a potential satisfying $V(\bm{q})=V(|\bm{q}|)$, we set $q_x=q\cos\theta$, $q_y=q\sin\theta$, $x=r\cos\phi$ and $y=r\sin\phi$, and find:
\bea
N(r,\phi)=
\cos(2\phi)\int_{+\infty}^0\frac{q{\rm d}q}{2\pi}\ph\frac{g}{c}\frac{q^2V(q)}{q^2+\xi_{\rm nem}^{-2}}J_2(qr)\,,\label{eq:angular_profile}
\eea

\noi with $J_2(z)$ the respective Bessel function of the first kind. One notes the distinctive angular dependence of the spatial profile of the induced nematic order, which is fixed by the B$_{\rm 1g}$ IR of $N$, the $A_{\rm 1g}$ IR of $V$, and the fourfold-symmetric impurity profile. We resolve the radial dependence in the case $V(\bm{r})=V/r$, and find:
\begin{align}
N(r,\phi)=\frac{\gamma}{c}\left[I_{2}\left(\frac{r}{\xi_{\rm nem}}\right)-L_{-2}\left(\frac{r}{\xi_{\rm nem}}\right)\right]\cos(2\phi),\label{eq:profile}
\end{align}

\noi where we introduced the modified Bessel and Struve functions, and defined $\gamma=-\pi gV/(2\xi_{\rm nem})$. Notably, the decaying function in the brac\-kets yields $\approx 1/2$ for $r=\xi_{\rm nem}$. 

\subsection{Non-Induction of Net Nematicity by a C$_4$-Symmetric Potential}

In this section we explore whether there exists a term in the Ginzburg-Landau expansion which can induce a nonzero $N(\bm{q}=\bm{0})$ for an impurity-potential profile which preserves C$_4$ symmetry. Consider the most general term:
\bea
\int {\rm d}\bm{r}\ph[N(\bm{r})]^{n}[V(\bm{r})]^m\big(\partial_x^2-\partial_y^2\big)^{\ell}V(\bm{r}) 
\eea

\noi where, if $\ell$ is odd, then $n=\ell+2\mathbb{N}$. We fix the spatial profile of $V$ to be C$_4$-symmetric. The above general term can be mapped to two distinct types of couplings:
\bea
\int {\rm d}\bm{r}\ph[N(\bm{r})]^{2n} \phd{\rm and}\phd \int {\rm d}\bm{r}\ph[N(\bm{r})]^{2n+1}\big(\partial_x^2-\partial_y^2\big)V(\bm{r})\,.
\eea

\noi The respective equations of motion read: 
\begin{align}
N(\bm{r})\propto [N(\bm{r})]^{2n-1}\ph{\rm and}\ph N(\bm{r})\propto [N(\bm{r})]^{2n}\big(\partial_x^2-\partial_y^2\big)V(\bm{r})\,.
\end{align}

\noi We Fourier transform the first equation and find:
\bea
&&N(\bm{q}=\bm{0})\propto\no\\&&\int {\rm d}\bm{p}_1\ldots {\rm d}\bm{p}_{2n-1} N(\bm{p}_1)\ldots N(\bm{p}_{2n-1})\delta\left(\sum_s^{2n-1}\bm{p}_s\right)\,.\qquad
\eea

\noi Assuming that the components appearing on the rhs are given by: 
\bea
\bar{N}(\bm{q})=\frac{g}{c}\frac{\big(q_x^2-q_y^2\big)V(\bm{q})}{\bm{q}^2+\xi_{\rm nem}^{-2}}\equiv
\cos(2\theta)\ph\frac{g}{c}\frac{q^2V(q)}{q^2+\xi_{\rm nem}^{-2}}\,,
\eea

\noi where we set $q_x=q\cos\theta$ and $q_y=q\sin\theta$, we find that the angular part of the integral is proportional to:
\bea
&&\int_0^{2\pi}{\rm d}\theta_1\ldots\int_0^{2\pi}{\rm d}\theta_{2n-2}\cos(2\theta_1)\ldots\cos(2\theta_{2n-2})\cdot\qquad\no\\
&&\left[\sum_{s=1}^{2n-2}p_s^2\cos\big(2\theta_s\big)+\sum_{s\neq\ell}^{2n-2}p_sp_\ell\cos\big(\theta_s+\theta_{\ell}\big)\right]=0\,,\qquad
\eea

\noi where we set $\cos\theta_s=p_{s,x}/p_s$ and $\sin\theta_s=p_{s,y}/p_s$, with $p_s=|\bm{p}_s|$. A similar treatment for the remaining equation also yields zero. This result naturally confirms, that, a C$_4$-symmetric spatial profile for the impurity potential cannot lead to net nematicity. 

\subsection{Case Study: Single Impurity for $\bm{T<T_{\rm nem}}$} 

In order to explain the elongated clover-like spatial profile induced by the impurity in the bulk nematic phase ($T< T_{\rm nem}$), we need to include higher order terms in the free energy described by Eq.~\eqref{eq:EOMfull} of the SM. To demonstrate how this elongation comes about, it is sufficient to solely retain the first term of Eq.~(5) presented in the main text. The modified EOM has the form:
\bea
&&\quad\big[\alpha(T-T_{\rm nem})-c\bm{\nabla}^2\big]N(\bm{r})+\beta[N(\bm{r})]^3\quad\no\\
&&\phd=-g\big(\partial_x^2-\partial_y^2\big)V(\bm{r})+g'N(\bm{r})V(\bm{r})\,.\qquad\qquad\label{eq:EOMmod}
\eea

\noi We separate the nematic field into two parts, i.e. $N(\bm{r})=N_{\rm B}+\delta N(\bm{r})$. Here, $N_{\rm B}$ denotes the value of the bulk nematic order parameter and is given by $\beta N_{\rm B}^2=\alpha(T_{\rm nem}-T)$ for $T<T_{\rm nem}$, while $\delta N(\bm{r})$ denotes the contribution stemming from the presence of the impurity. For $|\delta N(\bm{r})|\ll|N_{\rm B}|$ we linearize the above EOM and obtain:
\bea
&&\big[2\alpha(T_{\rm nem}-T)/c-\bm{\nabla}^2\big]\delta N(\bm{r})\qquad\no\\
&&\phd=-\frac{g}{c}\left(\partial_x^2-\partial_y^2\right)V(\bm{r})+\frac{g'}{c}N_{\rm B}V(\bm{r})\,.
\eea

\noi In the above, we retained the terms which lead to a $\delta N(\bm{r})$ which is linear in terms of the strength of the impurity potential. Within this assumption, we dropped the term $\delta N(\bm{r})V(\bm{r})$ which leads to higher-order contributions with respect to the strength of the impurity potential. In the same line of arguments as the ones leading to Eq.~\eqref{eq:angular_profile}, we obtain a constant angular profile superimposed on the usual $\cos(2\phi)$-form: 
\bea
\delta N(r,\phi)&=&\cos(2\phi)\int_{+\infty}^0\frac{q{\rm d}q}{2\pi}\ph\frac{g}{c}\frac{q^2V(q)}{q^2+\xi_{\rm nem}^{-2}}J_2(qr)\quad\no\\
&&\phd-N_{\rm B}\int^{0}_{+\infty}\frac{q{\rm d}q}{2\pi}\ph\frac{g'}{c}\frac{V(q)}{q^2 + \xi_{\rm nem}^{-2}}J_0(qr)\,,\quad
\eea

%

\noi with the coherence length being given now by $\xi_{\rm nem}^{-2}=2\alpha(T_{\rm nem}-T)$ due to the contribution of the quartic term of the free energy. In connection to Eq.~\eqref{eq:profile} of the SM, we find that for $V(\bm{r})=V/r$:
\bea
\delta N(r,\phi)&=&\frac{\gamma}{c}\left[I_{2}\left(\frac{r}{\xi_{\rm nem}}\right)-L_{-2}\left(\frac{r}{\xi_{\rm nem}}\right)\right]\cos(2\phi)\no\\
&-&\frac{\gamma'}{c}\left[I_{0}\left(\frac{r}{\xi_{\rm nem}}\right)-L_{0}\left(\frac{r}{\xi_{\rm nem}}\right)\right]N_{\rm B}\no\\
&\equiv&f(r)\cos(2\phi)+h(r)N_{\rm B}
\eea

\noi with $\gamma'=-\pi g'V\xi_{\rm nem}/2$. From the above, one can read off the decaying functions $f(r)$ and $h(r)$ discussed in the main text. This spatial profile does indeed lead to a profile on the same form as the anisotropic induced order in Fig.~2(b) of the main text. Furthermore, note that it is the presence of the nonzero bulk nematic order parameter $N_{\rm B}$, that induces the anisotropy. 


\section{Interaction in the Nematic Channel and Mean-Field Theory Decoupling}

We assume the presence of the interaction 
\bea
\widehat{\cal H}_{\rm int}=-V_{\rm nem}\sum_{\bm{R}}\widehat{{\cal O}}_{\bm{R}}^2/2\,,
\eea

\noi which contributes to the desired nematic channel. In the above, we have introduced: 
\bea
\widehat{{\cal O}}_{\bm{R}}=\sum_{\bm{\delta}}^{\pm\hat{\bm{x}},\pm\hat{\bm{y}}}f_{\bm{\delta}}
\left(c^{\dag}_{\bm{R}+\bm{\delta}}c_{\bm{R}}+c^{\dag}_{\bm{R}}c_{\bm{R}+\bm{\delta}}\right)\,,
\eea

\noi where we have defined the form factor $f_{\pm\hat{\bm{x}}}=-f_{\pm\hat{\bm{y}}}=1/4$. Note that the lattice constant has been set to unity. We perform a mean-field decoupling of the interaction in the direct channel by introducing the nematic order parameter $N_{\bm{R}}=-V_{\rm nem}\big<\widehat{{\cal O}}_{\bm{R}}\big>$. The latter steps led to Eq.~(9) of the main text.

In wavevector space we have $N_{\bm{q}}=\sum_{\bm{R}}N_{\bm{R}}e^{-i\bm{q}\cdot\bm{R}}$ and the complete mean-field Hamiltonian reads: 
\begin{align}
\widehat{{\cal H}}=\frac{1}{\cal N}\sum_{\bm{q},\bm{k}}c^{\dag}_{\bm{k}+\bm{q}/2}
\big(\varepsilon_{\bm{k}}{\cal N}\delta_{\bm{q},\bm{0}}+V_{\bm{q}}+N_{\bm{q}}f_{\bm{q},\bm{k}}\big)c_{\bm{k}-\bm{q}/2}
\end{align}

\noi with ${\cal N}$ being the number of lattice sites, while the nematic form factor in wavevector space takes the form:
\bea
f_{\bm{q},\bm{k}}=\frac{f_{\bm{k}+\bm{q}/2}+f_{\bm{k}-\bm{q}/2}}{2}\phd {\rm with}\phd f_{\bm{k}}=\cos k_x-\cos k_y\,.\ph
\eea

\noi The mean-field Hamiltonian has to be supplemented with the self-constistency equation for the nematic order parameter, which reads
\bea
N_{\bm{q}}
&=&-V_{\rm nem}\sum_{\bm{k}}f_{\bm{q},\bm{k}}\big<c^{\dag}_{\bm{k}-\bm{q}/2}c_{\bm{k}+\bm{q}/2}\big>\no\\
&\equiv&-V_{\rm nem}T\sum_{k_n,\bm{k}}f_{\bm{q},\bm{k}}G_{\bm{k}+\bm{q}/2,k_n;\bm{k}-\bm{q}/2,k_n}
\eea

\noi where we introduced the full single-particle fermionic Matsubara Green function:
\bea
G_{\bm{k}+\bm{q}/2,k_n;\bm{k}-\bm{q}/2,k_n}=-\big<c_{\bm{k}+\bm{q}/2,k_n}c^{\dag}_{\bm{k}-\bm{q}/2,k_n}\big>\,.
\eea

\noi In the above, $k_n=(2n+1)\pi T$ ($k_{\rm B}=1$) and the Matsubara Green function for the free electrons has the form $G^0_{\bm{k},k_n}=1/(ik_n-\varepsilon_{\bm{k}})$. The above construction allows us to employ Dyson's equation in order to perform an expansion of the rhs of the self-consistency equation with respect to the nematic order parameter and/or the impurity potential.

\section{Ginzburg-Landau Theory: Microscopic Analysis}

Given the above, here we show how the electro-nematic coefficient $g$ relates to the microscopic parameters for the disorder-free microscopic model under consideration. We employ a perturbative expansion by employing the Dyson equation for the full Matsubara Green function which reads:
\bea
&&G_{\bm{k}+\bm{q}/2,k_n;\bm{k}-\bm{q}/2,k_n}=G^0_{\bm{k},k_n}\delta_{\bm{q},\bm{0}}\no\\
&+&G^0_{\bm{k}+\bm{q}/2,k_n}\sum_{\bm{p}}U_{\bm{p};\bm{k}+\bm{q}/2}G_{\bm{k}+\bm{q}/2-\bm{p},k_n;\bm{k}-\bm{q}/2,k_n}\,,\qquad
\eea

\noi where we introduced $U_{\bm{q};\bm{k}}=\big(V_{\bm{q}}+N_{\bm{q}}f_{\bm{q},\bm{k}}\big)/{\cal N}$. We obtain the lowest order contribution of $U$ by replacing the full Green function on the rhs by the bare one. We find:
\bea
%
g_{\bm{q}}
=-\frac{T}{\cal N}\sum_{k_n,\bm{k}}f_{\bm{q},\bm{k}}G^0_{\bm{k}+\bm{q}/2,k_n}G^0_{\bm{k}-\bm{q}/2,k_n}\,.
\eea

\noi To facilitate the calculations, we consider the continuum limit of our model and assume spinless single-band electrons with a parabolic dispersion  $\varepsilon(\bm{k})=E_F\big[\left(k/k_F\right)^2-1\big]$ with $\bm{k}=(k_x,k_y)$, $k=|\bm{k}|$ and set $f(\bm{k})=\big(k_x^2-k_y^2\big)/k_F^2$. The quantity of interest, after taking into account the symmetries of $\varepsilon(\bm{k}),f(\bm{k})$ and restricting up to second order terms in $\bm{q}$, reads:

\bea
&&g(\bm{q})\approx
-\int \frac{{\rm d}\bm{k}}{(2\pi)^2}\bigg\{n_F'[\varepsilon(k)]\no\\
&&\quad+\big[f(\bm{k})\big]^2\frac{1}{3}E_F^2n_F'''[\varepsilon(k)]\bigg\}f(\bm{q}/2)
\equiv g\big(q_x^2-q_y^2\big)\,.\qquad
\eea

\section{Self-Consistent Calculation of the Nematic Order Parameter}

By means of the microscopic Hamiltonian in Eq.~(10) of the main text, we calculate the nematic order pa\-ra\-me\-ter self-consistently until we reach an accuracy of $10^{-6}$, while keeping the electron density fixed. The expectation values entering in the order parameter and the electron density are calculated by expressing the fermionic field operators in the diagonal basis of the Hamiltonian $c_{\bm{R}}=\sum_{m}\gamma_m\langle m|\bm{R}\rangle$ with the defining equation $\widehat{\cal H}\gamma^{\dagger}_m|0\rangle=E_m|m\rangle$. This leads to the following simplified expressions for the order parameter, and electron density, respectively:
\begin{align}
N_{\bm{R}}&=-V_{\rm nem}\sum_{\bm{\delta},\,m}f_{\bm{\delta}}\langle \bm{R}+\bm{\delta}|m\rangle n_{F}(E_m)\langle m|\bm{R}\rangle+{\rm c.c.}\,,\nonumber
\\
\langle n\rangle&=\frac{1}{\mathcal{N}}\sum_{m}n_{F}(E_m)\,.
\end{align}

\section{Disorder-Modified Stoner Criterion and the Resulting $\bm{T_{\rm nem}^{\rm imp}}$}

In the presence of dilute and uncorrelated identical impurities, the disorder may enhance the $T_{\rm nem}$. This was shown in the main text by investigating the modified nematic Stoner criterion. In Fig.~1 of the SM, we provide additional results for other electron-density values. The electron density is calculated via:
\begin{align}
\langle n\rangle=\frac{1}{\mathcal{N}}\sum_{\bm{k}}\int_{-\infty}^{\infty}\frac{{\rm d}\varepsilon}{2\pi}\frac{1}{\tau_{\bm{k}}}\frac{n_{F}(\varepsilon)}{(\varepsilon-\varepsilon_{\bm{k}})^2+1/(2\tau_{\bm{k}})^2},
\end{align}

\noi which recovers its usual form $\langle n\rangle=\sum_{\bm{k}}n_{F}(\varepsilon_{\bm{k}})/\mathcal{N}$ in the disorder-free case, i.e. $\tau_{\bm{k}}\rightarrow \infty$. For these calculations finite size effects are diminishing for $\mathcal{N} \sim 40\times 10^{3}$.

In Fig.~1 we demonstrate two typical situations, in which, $T_{\rm nem}$ becomes either enhanced or reduced. This is reflected in the behavior of the quantity $\delta\chi_{\rm nem}/\chi^0_{\rm nem}\equiv(\chi_{\rm nem}^{\rm imp}-\chi^0_{\rm nem})/\chi^0_{\rm nem}$ which is depicted. We first focus on $n_{\rm imp}$ in the vicinity of $5\%$, i.e. the optimal value discussed in the main text. 

For the value $\langle n\rangle=0.51$ of the electron density, the Fermi energy is tuned very near the van Hove singu\-la\-ri\-ty (see Figs.~1(a,b)), which constitutes the sweet spot for the development of the nematic order parameter in the absence of disorder, since there, $\chi^0_{\rm nem}$ obtains its ma\-xi\-mum value. From Fig.~1(c) we find that introducing disorder worsens the tendency of the system to develop a nematic order parameter as reflected in the negative va\-lues of $\delta\chi_{\rm nem}/\chi^0_{\rm nem}$. The addition of disorder broadens the spectral function, and the density of states (DOS) unavoidably becomes lowered, since contributions from low DOS $\bm{k}$ points are taken into account. In contrast, in the case $\langle n\rangle=0.53$ discussed in the main text, and also shown here, the broadening allows the DOS to increase by picking up contributions from the van Hove singularity, while at the same time avoiding significant contributions from other low DOS $\bm{k}$ points. Increasing the electron density to $\langle n\rangle=0.55$ shifts the Fermi level further away from the van Hove singularity and thus reduces its favorable impact on the DOS. As a result, the nematic susceptibility drops and $\delta\chi_{\rm nem}/\chi^0_{\rm nem}$ is negative. 

The balance between the contributions to the DOS originating from the van Hove singularity and the low DOS $\bm{k}$ points is controlled by the concentration of impurities. Varying $n_{\rm imp}$ leads to a modification of the re\-la\-ti\-ve strength of the two competing contributions and allows the sign changes of $\delta\chi_{\rm nem}/\chi^0_{\rm nem}$ which are shown in Fig.~1(c) for $\langle n\rangle=0.55$. 

\begin{figure}[b!]
\centering
\includegraphics[width = \columnwidth]{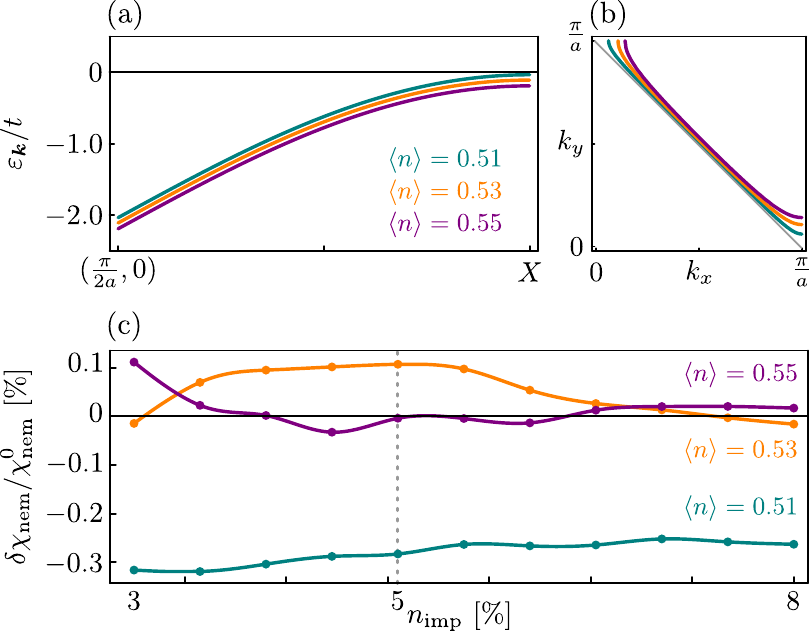}
\caption{(a) Energy dispersion along the ${\rm \Gamma}-{\rm X}$ line, (b) Fermi line in $(k_x,k_y)$ space and (c) $\delta\chi_{\rm nem}/\chi^0_{\rm nem}$ as a function of $n_{\rm imp}$, all shown for different electron fillings $\langle n\rangle=0.51,0.53,0.55$. Panel (c) reveals that disorder always has a negative impact on the nematic susceptibility when the Fermi level is tuned very near the van Hove singularity, as inferred for $\langle n\rangle=0.51$. When the Fermi level is tuned suffieciently away from the van Hove singularity, the resulting nematic susceptibility can be either enhanced or reduced depending on the relative strength of the contributions to the density of states (DOS) stemming from the van Hove singularity and the low DOS $\bm{k}$ points. This ratio is controlled by the concentration of impurities $n_{\rm imp}$.}
\end{figure}

\end{document}